# Lithium Mediated Benzene Adsorption on Graphene and Graphene Nanoribbons

Dana Krepel and Oded Hod

School of Chemistry, The Raymond and Beverly Sackler Faculty of Exact Sciences, Tel Aviv University, Tel Aviv 69978, Israel

Abstract:

The anchoring of benzene molecules on lithium adsorption sites at the surface of graphene and nanoribbons thereof are investigated. The effects of adsorbate densities, specific adsorption locations, and spin states on the structural stability and electronic properties of the underlying graphene derivatives are revealed. At sufficiently high densities, bare lithium adsorption turns armchair graphene nanoribbons metallic and their zigzag counterparts half-metallic due to charge transfer from the adatom to the $\pi$ electronic system. Upon benzene adsorption, the lithium cation encounters two $\pi$ systems thus drawing charge back towards the benzene molecule. This, in turn, leads to the opening of a measurable bandgap, whose size and character are sensitive to the adsorbate density, thus indicating that a chemical detector based on lithium adsorbed graphene may be devised. Our results are therefore expected to support the design of novel graphene-based sensing and switching devices.

**Introduction**

The successful isolation of graphene as a stand-alone two-dimensional all-carbon network[1] has attracted enormous attention over the past few years. Graphene nanoribbons (GNRs), which can be viewed as elongated stripes cut out of a graphene sheet, possess a variety of interesting physical characteristics such as ballistic electronic transport, current density sustainability,[2] quasi-relativistic behavior,[3] and electronic structure engineering capabilities.[4] Due to their unique structural, mechanical, and electronic properties, GNRs have been identified as important candidates for numerous potential applications including spintronic devices,[5] gas sensors,[6] and nano-composites.[6a, 7]

The interactions of graphene with different chemical species are one of the main ingredients for many of the potential applications of graphene-based devices. Most of the previous studies in this field have focused on the physisorption of small gas molecules on both pristine[8] and defected graphene[6b, 9] as well as carbon nanotubes.[10] Recently, non-covalent functionalization of graphene using various molecules of increasing complexity have attracted growing attention. It has been reported that several organic molecules[6c, 11] or even bio-molecules[12] could induce significant changes in the electronic properties of graphene. Introducing dopants such as metal atoms has been reported to enhance the molecule/graphene interaction, which is highly important for sensing devices.[9a, 13] Among all alkali-metals, lithium, having the smallest atomic radius, has been identified as the most strongly binding atom to the surface of pristine two-dimensional graphene[14] due to strong cation-π interactions.[15] Lithium adsorption on graphene derivatives of reduced dimensionality such as armchair (AGNRs) and zigzag (ZGNRs) graphene nanoribbons was further studied showing that the favorable adsorption sites on the nanoribbons are the hollow sites near the edges of the ribbon.[16] In addition, adsorption of the metal atoms on semi-conducting GNRs was found to significantly decrease the bandgaps of the ribbons, turning them metallic for sufficiently high adatom densities.[16b]

In the present work, we report the results of first-principles calculations on the lithium mediated adsorption of benzene, the simplest aromatic molecule, which serves as one of the most common intermediate species for various chemical compounds, on the surface of two-dimensional (2D) graphene and both AGNRs and ZGNRs. This unique adsorption scheme facilitates a π-metal-π bridging interaction, which is known to be stronger than direct π-π dispersive interactions [11c]. Hence, the purpose of this study is to gain insight on the π-metal-π sandwich-like structure and to examine the effects of the adsorbed complex on the relative stability and the electronic properties of graphene and graphene nanoribbons, taking into account different ribbon widths, adsorbate densities, and spin states.

**Methods**

Our density functional theory (DFT) calculations have been carried out utilizing the Gaussian suite of programs.[17] Spin-polarized calculations have been performed within the generalized gradient approximation of Perdew et al. (PBE),[18] and the screened-exchange hybrid functional of Heyd, Scuseria, and Ernzerhof (HSE).[19] The latter has been shown to reproduce experimental optical bandgaps of bulk semi-conductors and to describe the physical properties of graphene-based materials with much success.[4b, 20] We use the double-zeta polarized 6-31G$^{**}$ basis set[16,21] noting that the effects of basis set superposition errors (BSSE)[22] in this type of systems have been recently studied in detail and shown to be of minor importance at this level of theory.[16]

**Results and Discussion**

Lithium mediated benzene adsorption on two-dimensional graphene

We start by revisiting the problem of benzene adsorption on 2D graphene sheets. It is known that benzene binds more strongly with metal/graphene than with pristine graphene.[11d, 23] Therefore, we designed a graphene/metal-atom/molecule sandwich structure. First, lithium atoms were placed at their preferred graphene adsorption cites on top of hexagon centers. To form the sandwich structure, benzene molecules were then situated on the lithium-graphene structure, with their centers located above the lithium atoms. Previous studies suggested that the ground state of the bare-lithium-graphene system is of singlet closed-shell nature.[16b] In order to examine consistency with these results for the full sandwich structure, we performed calculations for three unit-cells with increasing dimensions representing decreasing dopant densities. We annotate the unit-cells by ($N$x$M$) where $N$ stands for the number of zigzag chains along the armchair edge and $M$ for the number of carbon dimers along the zigzag edge.[4c] Using this notation we choose the (4×8), (6×10) and (6×12) supercells with a single lithium-benzene complex calculated at the doublet spin state (panels a–c of Fig.1). The singlet and triplet states are then obtained by duplicating the unit-cells along the periodic armchair direction, creating (8×8), (12×10) and (12×12) supercells (panels d–f in Fig.1). For these spin states, we consider two Li-benzene complexes located on the opposite sides of the graphene sheet, that is, one above the center of a hexagon of the graphene sheet, and the other below the center of another hexagon distant from the first one. Evidently, this minimizes the interactions between the two partially positively charged metal atoms[11c] as well as the steric repulsions between the benzene molecules.

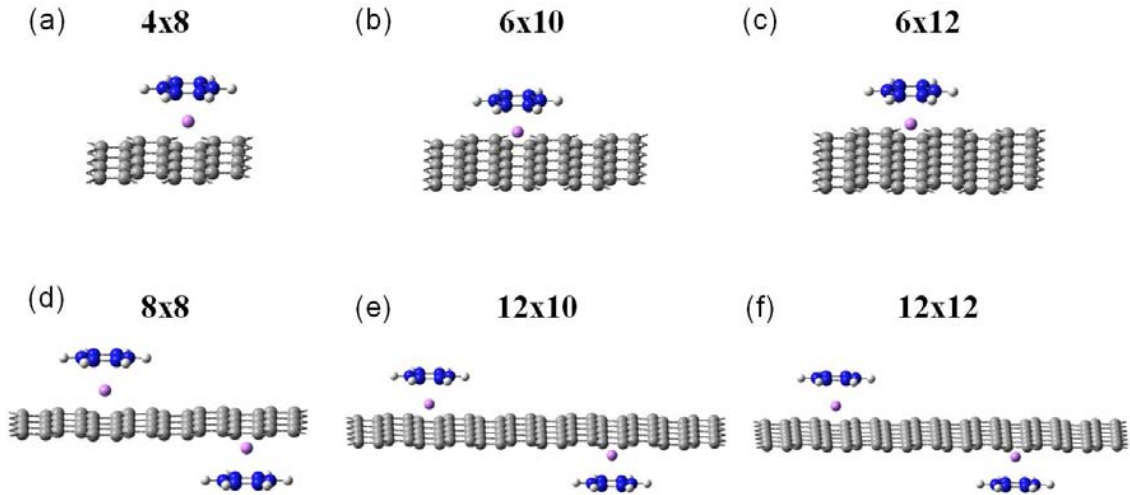

Figure 1: A schematic representation of the (a) 4×8, (b) 6×10 and (c) 6×12 2D graphene supercells and their (d) 8×8, (e) 12×10 and (f) 12×12 counterparts. Gray, white, and magenta spheres represent carbon, hydrogen, and lithium atoms, respectively. The carbon atoms of the benzene molecules have been marked in blue for clarity of the representation.

As can be seen in Fig. 2, and consistent with the adsorption of bare lithium atoms on 2D graphene, for all unit-cells considered the closed-shell singlet state was found to be the most stable spin state lower by 0.20–0.35 eV/Li-benzene complex than the corresponding doublet and triplet states for both functional approximations considered.

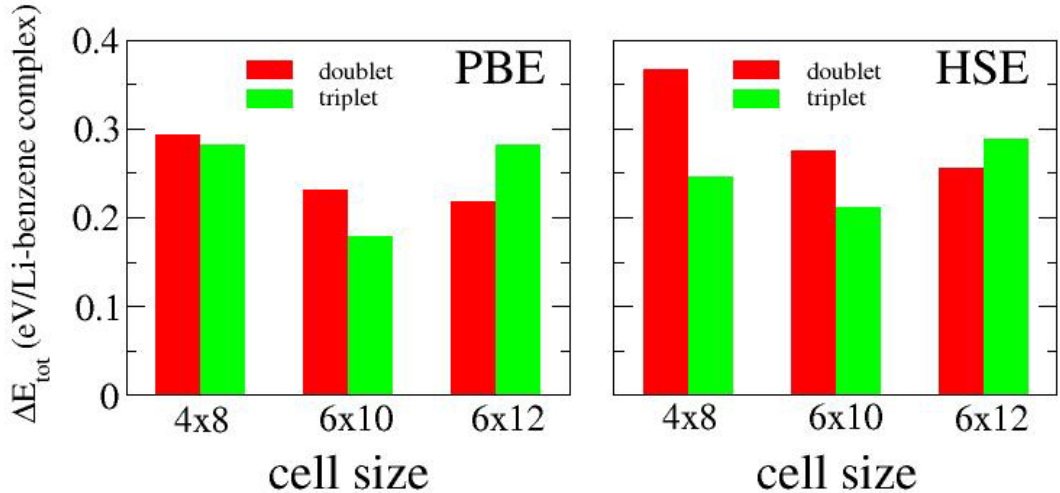

Figure 2: Total energies of the doublet and triplet spin states with respect to the closed-shell singlet spin state for the three 2D graphene unit-cell sizes studied as calculated using the PBE (left panel) and HSE (right panel) functional approximations.

In order to further examine the influence of different dopant densities and spin states we have performed binding energy calculations. The upper panels of Fig. 3 present full binding energy curves of the 4x8, 6x10, and 6x12 unit-cells as calculated using the PBE (upper left panel) and HSE (upper right panel) density functional approximations. The obtained PBE binding energy for the 6x10 unit-cell was found to be ~1.00 eV/Li-benzene complex which is consistent with previous calculations yielding 0.97 eV/unit-cell for this system.[11d] The corresponding HSE binding energy was ~1.05 eV/Li-benzene complex. When comparing the binding energies obtained for the different unit-cell sizes considered, we find that lower dopant densities result in somewhat (~0.05-0.1 eV/Li-benzene complex) stronger binding. This may be rationalized by the larger distance and hence reduced interactions between lithium-benzene complexes residing on adjacent unit-cells.

The lower panels of Fig. 3 present the HSE binding energy curves for the 4x8 (lower left panel) and 6x10 (lower right panel) unit-cells calculated at different spin states. For both systems the closed-shell singlet state shows a higher binding energy by ~0.1 eV/Li-benzene complex as compared to both the doublet and triplet spin states, thus further demonstrating that the ground state for this system is of closed-hell singlet spin nature.

It is important to mention that our calculations do not include long-range correlation effects and thus neglect van der Waals interactions. Nevertheless, for the binding scheme considered herein, dispersion interactions are expected to be less important than the cation-π interactions and we expect their inclusion to somewhat increase the calculated binding energies. Therefore, our results may be viewed as a lower bound for the true binding energies.

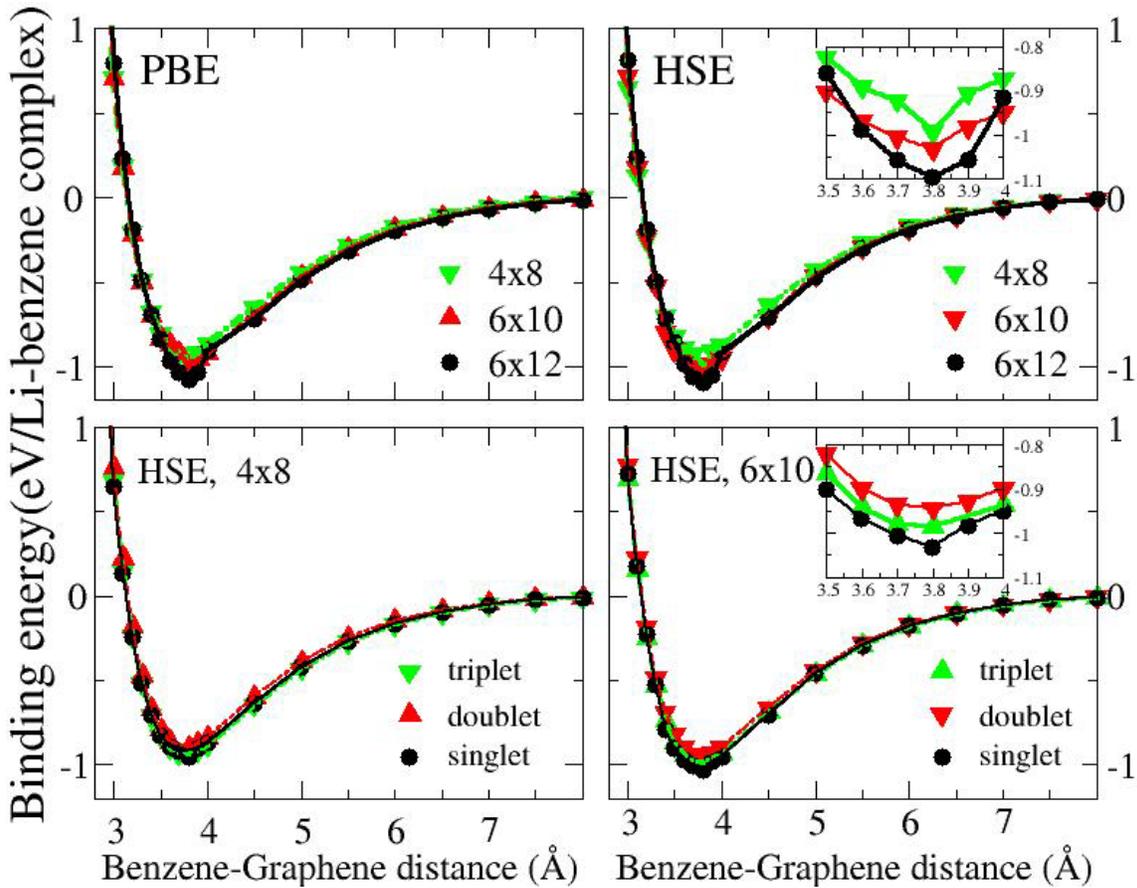

Figure 3: Binding energy of benzene on lithium doped 2D graphene at the closed-shell singlet spin state calculated for the (4×8), (6×10) and (6×12) unit-cells using the PBE (upper left panel), and HSE (upper right panel) functionals. The inset of the upper right panel presents a zoom-in on the binding energy curves for the various systems considered near the equilibrium distance. Lower panels: A comparison of the binding energies calculated for the closed-shell singlet, doublet and triplet spin states using the HSE functional for the 4x8 (left) and 6x10 (right) unit-cells. The inset of the lower right panel presents a zoom-in on the binding energy curves around the equilibrium distances of the different spin states considered.

An important aspect of the graphene/lithium/benzene sandwich structure is the charge redistribution upon benzene adsorption. Once a benzene molecule is adsorbed, electrons are withdrawn from the graphene π system towards the benzene molecule, thus altering the electronic doping of the graphene. In order to characterize these effects we plot in Fig. 4 the charge density differences defined as:

$$\delta\rho = \rho(grahpene + Li + benzene) - [\rho(graphene + Li) + \rho(benzene)], \quad (1)$$

where $\rho(grahpene + Li + benzene)$ is the electron density of the graphene/lithium/benzene complex, $\rho(graphene + Li)$ is the electron density of the graphene/lithium complex, and $\rho(benzene)$ is the electron density of the isolated benzene molecule, all calculated at the same level of theory.

We plot the charge density differences at four different benzene-graphene separations representing short distances (2.5 Å), slightly below (3.5 Å) and slightly above (4.0 Å) the equilibrium distance (~3.8 Å), and at a large separation of 6.0 Å. The results clearly show that at the larger distance, the systems become separated and no charge transfer occurs between the lithium atom and the benzene molecule (Fig 4(d)). As the distance decreases, charge transfer towards the benzene molecule increases (Fig. 4, panels (b) and (c), purple isosurfaces) due to enhanced cation-π interactions between the lithium cation and the π electronic system of the benzene molecule. Despite this charge transfer, it is important to note that the quasi-two-dimensional system remains metallic. When the benzene molecule is forced upon the lithium atom below the equilibrium distance, strong charge transfer towards the benzene molecule occurs which suppresses the transfer of charge to the 2D graphene surface (Fig. 4(a)).

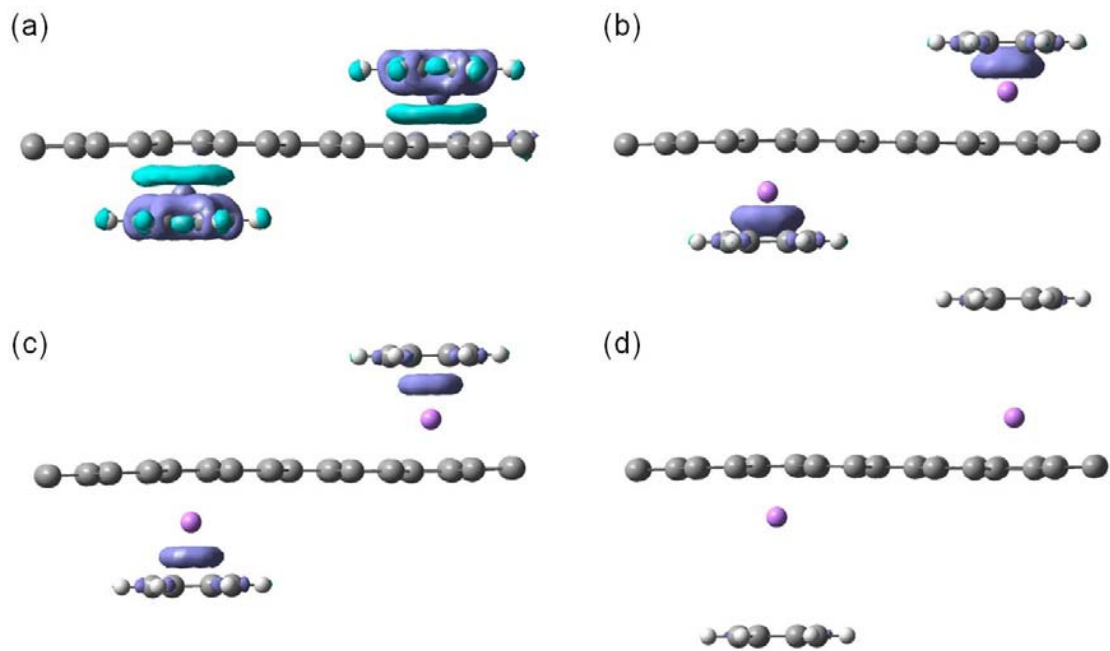

Figure 4: Charge density differences calculated using Eq. (1) for the two-dimensional 12×12 unit-cell at the HSE/6-31G** level of theory obtained at benzene-graphene separations of: (a) 2.5 Å, (b) 3.5 Å, (c) 4.0 Å and (d) 6.0 Å. Turquoise (purple) isosurfaces represent reduction (increase) in electron density with respect to the separated systems.

Lithium mediated benzene adsorption on quasi-one-dimensional graphene nanoribbons

Having examined the interactions of lithium adatoms with 2D graphene, we are now in position to discuss their adsorption on the surface and edges of quasi-one-dimensional graphene nanoribbons. The ribbons considered in the present study are obtained by cutting the 2D graphene sheet along either its armchair or zigzag crystalline orientations while passivating the bare edges with hydrogen atoms.

For the armchair systems, we consider ribbons of three consecutive widths (N×7), (N×9), and (N×11) to represent the three subsets of AGNRs with varying bandgaps.[4a-d] For each ribbon width we study two unit-cell lengths of N=4 and N=6 giving a total set of six different systems. These represent adsorbate densities of 2 lithium-benzene complexes per $0.629$ nm$^2$, $0.838$ nm$^2$, and $1.048$ nm$^2$ for the shorter cells of increasing width, and 2 lithium-benzene complexes per $0.943$ nm$^2$, $1.258$ nm$^2$, and $1.572$ nm$^2$, for the longer cells, respectively (see Fig. 5).

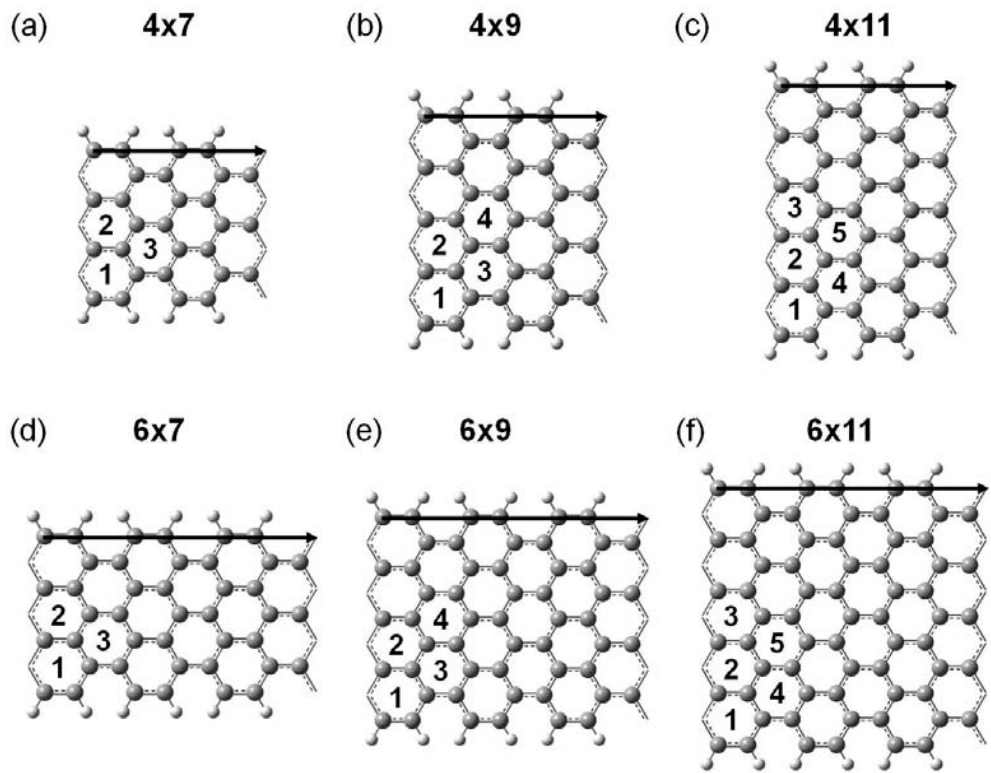

Figure 5: Schematic diagrams of the AGNRs supercells used for the (a) 4x7, (b) 4x9, (c) 4x11, (d) 6x7, (e) 6x9, and (f) 6x11 calculations. Various adsorption hollow sites (above hexagon centers) are marked with consecutive numbers. Arrows represent the translational vectors along the periodic direction.

For the ZGNRs, we study two consecutive widths of (6×M), and (8×M). For each ribbon width we consider two unit-cell lengths of M=6 and M=8 giving a total set of four different systems. These represent adsorbate densities of 2 lithium-benzene complexes per 0.576 nm$^2$ and 0.891 nm$^2$ for the shorter cells of increasing width, and 2 lithium-benzene complexes per 0.768 nm$^2$ and 1.188 nm$^2$ for the longer cells, respectively (see Fig. 6).

The adsorption of a single lithium-benzene complex and two such complexes per unit-cell are considered. In addition, for the single complex per unit-cell adsorption scenarios duplications of each cell along the periodic direction were performed to examine the effects of different spin states.

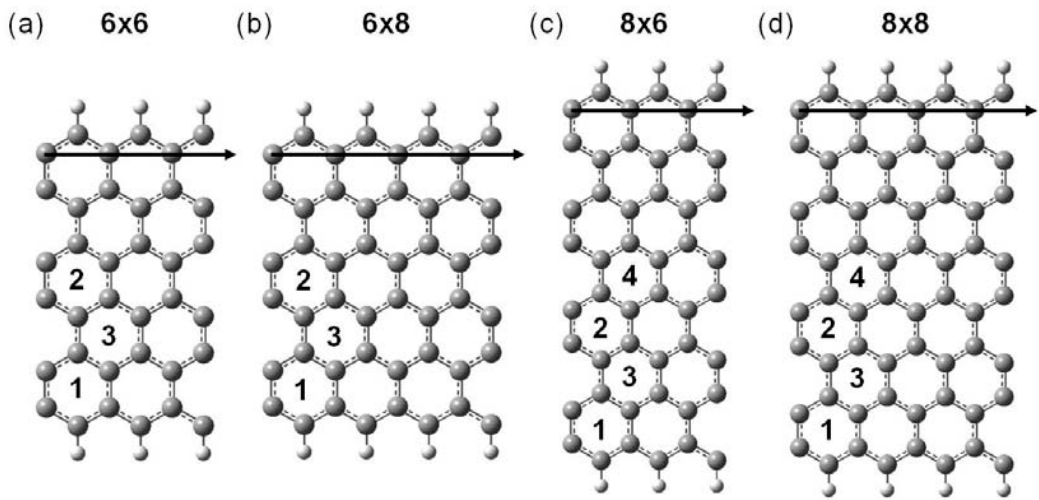

Figure 6: Schematic diagrams of the ZGNRs supercells used for the (a) 6x6, (b) 6x8, (c) 8x6, and (d) 8x8 calculations. Various adsorption hollow sites (above hexagon centers) are marked with consecutive numbers. Arrows represent the translational vectors along the periodic direction.

First, we study the relative stabilities of the different adsorption sites of the various GNRs considered. As mentioned above, the hollow sites are the most stable lithium adsorption positions on both AGNRs and ZGNRs.[16] Therefore, we systematically place a single lithium-benzene complex on top of all distinct hollow sites within the unit-cells considered (see site numberings in Figs. 5 and 6). Fig. 7 compares the total energies of the fully relaxed systems on different adsorption sites at the corresponding equilibrium distances obtained for the AGNRs and ZGNRs studied. Similar to the case of lithium atom adsorption,[16] we find that the most stable lithium-benzene complex adsorption sites are above the center of the hexagons close to the edges of the ribbon (position 1 in panels (a)–(f) of Fig. 5 and panels (a)-(d) of Fig. 6). As previously suggested, this behavior may be attributed to the reactive nature of the honeycomb lattice edges.[16, 24]

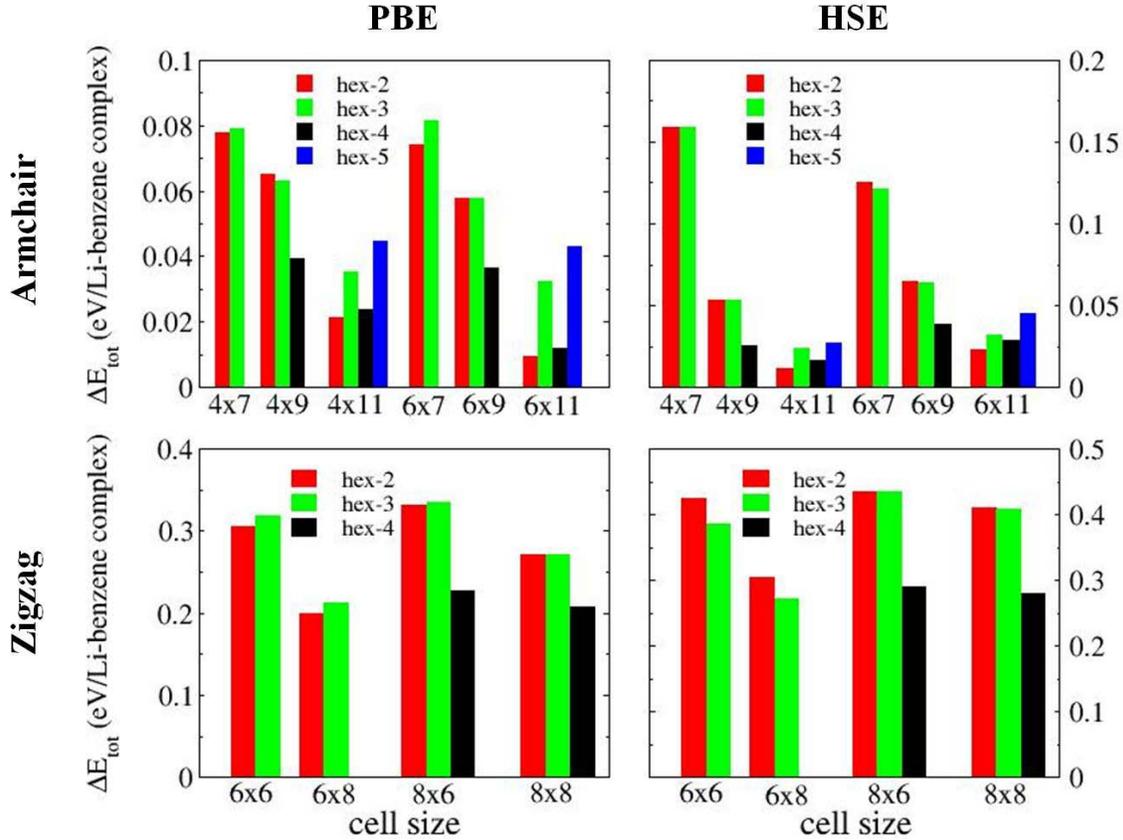

Figure 7: Closed-shell singlet relative energies of different adsorption sites for the six AGNRs unit-cell sizes (upper panels) and triplet relative energies of different adsorption sites for the four ZGNRs unit-cell sizes (lower panels) studied as calculated using the PBE (left panels), and HSE (right panels) functional approximations. The energies of the systems where adsorption occurs on the edge hollow site (position 1) are taken as a reference.

In order to study the effect of different spin states we duplicate the unit-cells and compare the results of the doublet spin state obtained with a single lithium-benzene complex per unit-cell at the edge adsorption site (position 1) to the singlet and triplet states calculated with the duplicated unit-cells. We note that for the ZGNRs, different spin configurations of the singlet state were taken into account, including an open-shell singlet anti-ferromagnetic (AFM) state, with anti-parallel spin orientation at the two zigzag edges, and the closed-shell singlet non-magnetic (NM) state. Representative spin orderings of each open-shell spin state are presented in Fig. 8.

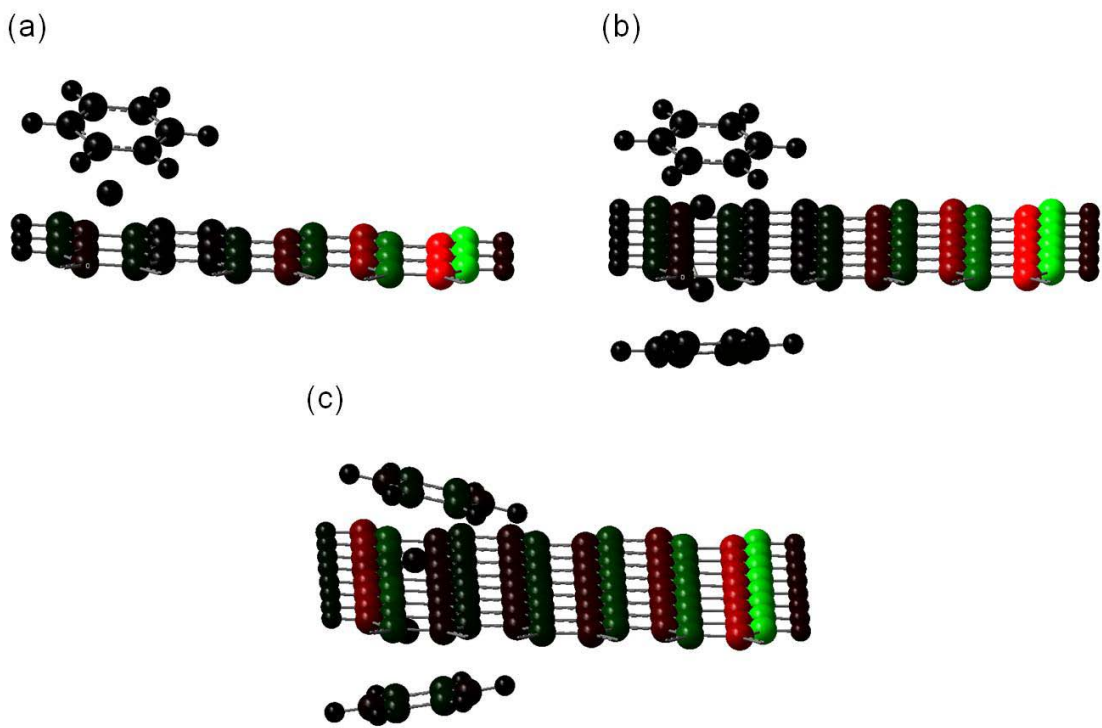

Figure 8: Representative spin configuration of the lithium-benzene adsorbed 6x8 ZGNR obtained for (a) Doublet spin state, (b) Triplet spin state and (c) Open-shell singlet AFM ordering as calculated at the HSE/6-31G** level of theory. Mulliken spin charge range is ±0.3 in all panels.

For all AGNRs studied we find that the closed-shell singlet spin state is the most stable state lower in energy by up to 0.32 eV per lithium-benzene complex than the corresponding doublet and triplet states (see upper panels of Fig. 9). Consistent with our findings for lithium adsorption on AGNRs,[16b] even at the largest unit-cell dimensions considered the singlet/doublet and singlet/triplet differences continue to increase thus indicating the long-range nature of the graphene-mediated lithium–lithium interactions in the GNRs systems studied. Unlike lithium-benzene complex adsorption on AGNRs, for all ZGNRs considered, we find that the triplet spin state is of lowest total energy (see middle panels of Fig. 9). The doublet spin state is found to be somewhat higher in energy than the triplet state with HSE total energy differences ranging from 0.008 to 0.12 eV/Li-benzene complex for the various ZGNRs considered. Significant triplet/singlet differences reaching ~0.37 eV/Li-benzene complex for the largest unit-cell studied are found indicating that the singlet spin-states are unfavorable for this system.

Here, it should be noticed that previous studies on the adsorption of bare lithium atoms on the surface and edges of ZGNRs have focused on the adsorption of a single adatom per cell-unit at the doublet spin state.[16a] In light of our findings above for lithium-benzene complexes, we have further examined the effect of different spin states in the case of bare lithium adsorption on the ZGNRs we study. In the lower panels of Fig. 9 we present the relative stability of the various spin states considered. Indeed, in the case of relatively small unit-cells the doublet and triplet spin states are of equivalent energetic stability and are considerably more stable than the singlet states. Nevertheless, as the adsorbate density decreases the triplet spin state becomes the most stable one with the doublet spin state lying ~0.05 eV/Li atom above it for the largest ZGNR considered. Therefore, consistent with our findings for the lithium-benzene complex adsorption, we conclude that for both lithium atom and lithium-benzene complex adsorption on ZGNRs, the favorable spin state is of triplet nature.

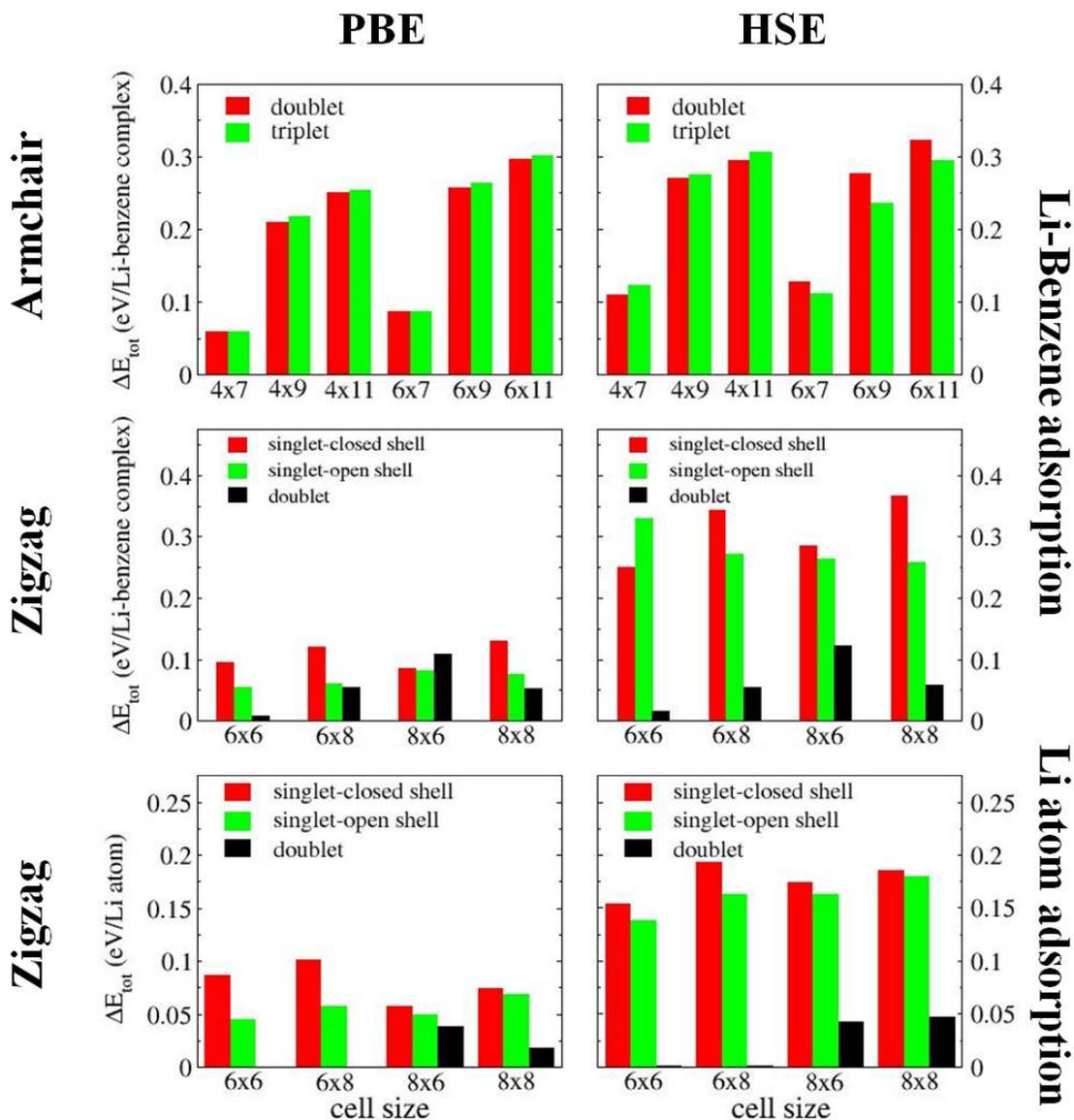

Figure 9: Effect of spin state on the energetic stability of the edge adsorption schemes considered. Upper panels: relative energies of the doublet and triplet spin states with respect to the closed-shell singlet spin state of the six lithium-benzene adsorbed AGNRs unit-cells considered as calculated using the PBE (upper left panel) and HSE (upper right panel) functional approximations. Middle panels: relative energies of the singlet and the doublet spin states with respect to the triplet state of the four Li-benzene adsorbed ZGNR unit-cells considered as calculated using the PBE (middle left panel) and HSE (middle right panel) functional approximations. Lower panels: relative energies of the singlet and the doublet spin states with respect to the triplet state of the four bare lithium atoms adsorbed ZGNR unit-cells considered as calculated using the PBE (lower left panel) and HSE (lower right panel) functional approximations.

To finalize the structural stability analysis we consider the adsorption of two Li-benzene complexes at various relative positions along the width of the ribbon's unit-cell. This enables us to further explore the relative stability of different complex adsorption sites and dopant densities. To this end, we place a single complex on top of the hollow site near one edge of the ribbon (positions 1' in Fig. 10), and systematically compare the adsorption of a second complex positioned below various possible hollow sites along the width of the ribbon (see unprimed positions in Fig. 10(b) and (d)). Since coulomb repulsion between the adsorbates is a major consideration in these systems we focus on relatively large unit-cells for these calculations, namely the 12x11 AGNR and the 8x12 ZGNR.

The total PBE and HSE energies of the different adsorption sites of the second complex are presented in Fig. 10(a) for the 12x11 AGNR at the closed-shell singlet spin state and 10(c) for the 8x12 ZGNR at the triplet spin state. Interestingly, the most stable configuration considered is obtained when the two complexes are located above hollow sites of hexagons near the same edge of the ribbon (adsorption sites 1' and 1 in panels (b) and (d) of Fig. 10). The next energetically favorable configuration is obtained when the two complexes are located near the opposite edges of both the AGNR (adsorption sites 1' and 5 in panel 10(b)) and ZGNR (adsorption sites 1' and 4 in panel 10(d)).[25] The least favorable metastable adsorption site of the second complex on the surface of the AGNR (position 6 in Fig. 10(b)) is found to be adjacent to the most favorable site (position 1). For the ZGNR the least favorable metastable adsorption site of the second complex is at the center of the ribbon (position 6 in Fig. 10(d)).

The inter-adsorbate interactions are more pronounced in case of the 8x12 ZGNR than the 12x11 AGNR, reaching a ~0.52 eV difference between the most stable and most unstable hollow adsorption sites for the HSE functional approximation. This may be attributed to the smaller supercell used to represent the ZGNR studied resulting in enhanced interactions between the Li-benzene adsorbates.

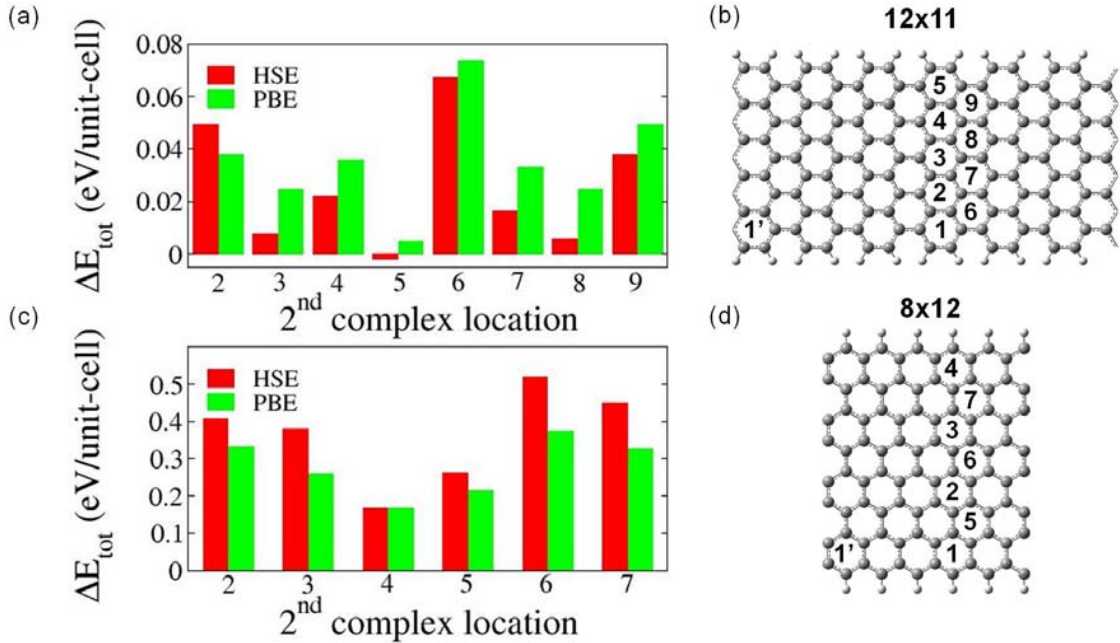

Figure 10: Relative total energies of the adsorption of two lithium-benzene complexes located at different adsorption sites along the (a) 12x11 AGNR and (c) 8x12 ZGNR calculated using the PBE (green bars), and HSE (red bars) functional approximations. The closed-shell spin state is used for the AGNR and the triplet state is used for the ZGNR. Values are presented with respect to the case where the two complexes are located near the same edge of the ribbon (positions 1' and 1 in panels (b) and (d)). Numbering of the various adsorption sites considered are marked in the schematic diagrams presented in panels (b) and (d) for the 12x11 AGNR and 8x12 ZGNR supercells, respectively.

After studying the structural stability of benzene molecules adsorbed on the surface and edges of lithium doped GNRs we now turn to examine their influence on the electronic properties of the underlying ribbon. Based on our findings above, we consider the closed-shell spin state for AGNRs and the triplet state for the doped ZGNRs and adsorb two Li-benzene complexes at their favorable adsorption sites along one edge of each super-cell studied. We compare the bandgaps of four types of systems (see Fig. 11): (a) the pristine GNRs; (b) lithium atoms adsorption along one edges of the GNR above and below the surface in an alternating pattern; (c) partial benzene adsorption, i.e. only atop lithium sites on one side of the GNR surface, leaving the second Li site bare; and (d) benzene adsorption on both lithium sites above and below the basal plane of the GNR.

Importantly, using the triplet spin state for the ZGNRs, we examine the influence of the different adsorption schemes on both α and β spin states.

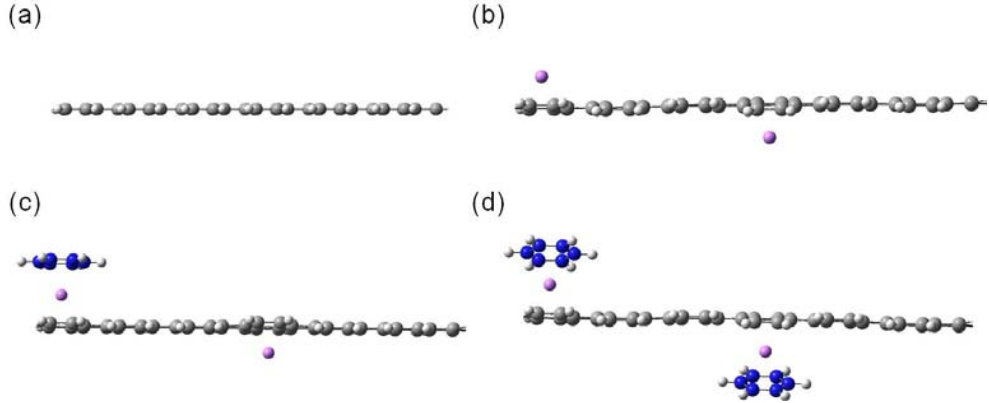

Figure 11: Schematic diagram of the various adsorption schemes on the duplicated 6x7 AGNR unit-cell used to study the changes in electronic structure of the ribbon: (a) The pristine nanoribbon; (b) Lithium atom adsorption along one edge of the ribbon in an alternating pattern above and below the surface; (c) Partial benzene adsorption on lithium anchoring sites only on one side of the surface; and (d) Benzene adsorption on both lithium sites above and below the surface. Gray, white, and magenta spheres represent carbon, hydrogen, and lithium atoms, respectively. The carbon atoms of the benzene molecules have been marked in blue for clarity of the representation.

In Fig. 12 we present the bandgaps obtained for the adsorption schemes depicted in Fig. 11 of the various AGNRs studied. Since the systems considered represent the three sub-families of AGNRs, the bandgaps of the pristine systems are found to considerably vary with the ribbon's width ranging from 0.2 eV to 2 eV as calculated using the HSE functional approximation.[4b] Upon lithium adsorption a dramatic decrease of the bandgap is evident with values not exceeding 0.05 eV and relatively weak dependence on the width of the ribbon.[16b] The effect of benzene adsorption on the Li anchoring sites both at the lower and the higher densities considered is clearly reflected in the bandgap of the system giving rise to increased values with respect to the corresponding lithium adsorbed AGNR. For all AGNRs considered, the increase was in the range of 0.06-0.07 eV for the HSE functional and 0.025-0.04 eV for the PBE functional. It should be noted that, as expected, the energy gaps obtained with the HSE functional are consistently larger than the corresponding PBE values. However, both functional approximations predict the same general trends. These results suggest that by measuring the electronic properties or transport characteristics of lithium doped AGNRs one should be able to identify adsorption events of benzene molecules at appropriate contaminant densities.

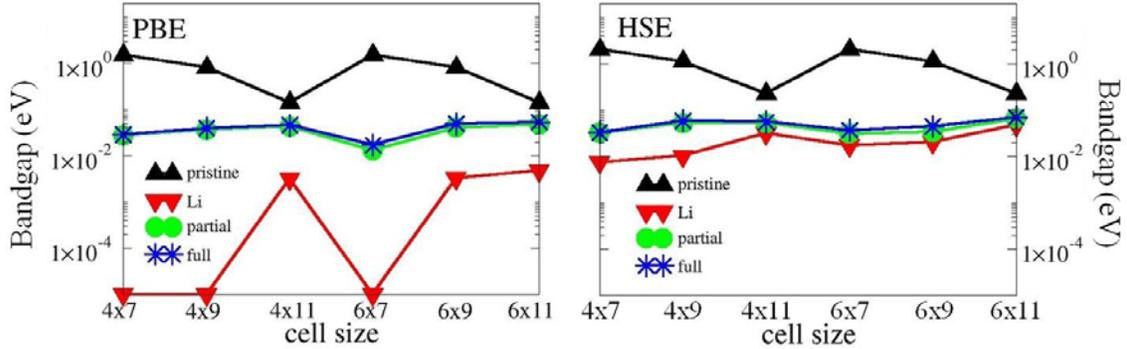

Figure 12: The influence of benzene adsorption on the bandgap of lithium doped AGNRs. Presented is the dependence of the closed-shell spin state bandgap on the unit-cell dimensions for hydrogen-terminated AGNRs after lithium adsorption (red downward facing triangles), partial Li-benzene adsorption (green circles) and full Li-benzene adsorption (blue stars) as calculated using the PBE (left panel) and HSE (right panel) density functional approximations. Results for the pristine nanoribbons appear as upward facing black triangles. Notice that a logarithmic scale is used to represent the bandgap values of the different systems.

In order to obtain a more complete picture of the effects of the various adsorption schemes on the electronic properties of the AGNRs studied we plot, in Fig. 13, the band-structures,[26] density of states (DOS),[27] and partial density of states (PDOS) obtained for the doped 6x7 duplicated AGNR unit-cell. As can be seen, the overall electronic structure changes considerably when comparing the pristine system to the lithium and lithium-benzene adsorbed systems. The wide direct bandgap appearing at the $\Gamma$ point of the pristine AGNR is replaced by a narrow direct X point gap of the chemically modified systems. When comparing the various adsorption schemes it is found that the qualitative structure of the bands near the Fermi energy is preserved. This can be rationalized by the fact that the major contribution to the DOS in the vicinity of the Fermi energy comes from the graphene backbone whereas the lithium and benzene contributions appear near 1-1.25 eV above the Fermi energy (see PDOS analysis). When focusing on the energy range close to the Fermi energy (see Fig. 13(e)) one finds that benzene adsorption mainly induces a relative up-shift of the valence band maximum (VBM) and a corresponding relative down-shift of the conduction band minimum (CBM) near the X-point. A similar picture is found for the other AGNRs studied (not shown).

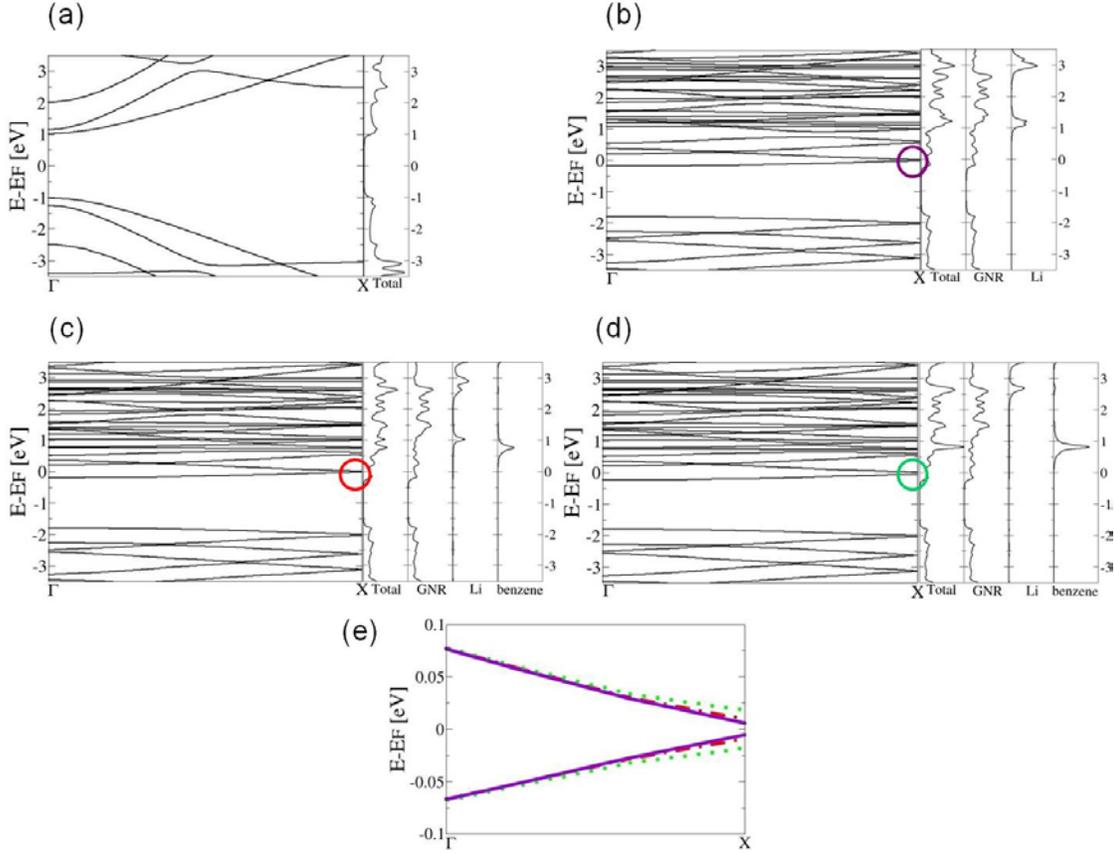

Figure 13: Closed-shell band-structures, DOS and PDOS of the duplicated 6x7 AGNR unit-cell at the (a) pristine form; (b) Lithium adsorbed scheme; (c) Partial benzene adsorption scheme; and (d) Full benzene adsorption scheme (see Fig. 11). Zoom in on an energy range of ±0.1eV around the Fermi energy of lithium adsorbed (full purple line), and partial (dashed red line) and full (green dotted line) lithium-benzene adsorbed 6x7 AGNR as calculated at the HSE/6-31G** level of theory is presented in panel (e). Fermi energies of all diagrams are set to zero.

In Fig. 14 we present results for the case of ZGNRs where a different qualitative behavior appears showing strong spin sensitivity towards the adsorptions of the lithium-benzene complexes. As is well established for pristine ZGNRs, the ground state is of singlet open-shell nature where anti-parallel spin ordering appears at the zigzag edges and equal bandgaps are obtained for both spin flavors.[28] Upon bare lithium atoms adsorption the bandgap degeneracy of the two spin flavors is lifted resulting in α HSE bandgaps in the range of 0-0.02 eV and β bandgaps reaching as high as 0.8 eV. This state resembles the half-metallic state previously reported for ZGNRs under the influence of an appropriately oriented external electric-field[20a, 28] as well as GNRs with various zigzag edge chemical terminations.[29] Further adsorption of the benzene molecules on the lithium anchoring sites of all systems considered, both at the lower and the higher adsorption densities, somewhat reduces the α/β bandgap split. The obtained α HSE bandgaps in this case are in the range of 0.05-0.10 eV and the corresponding β bandgaps are 045-0.74 eV. As may be expected, the corresponding PBE values are consistently lower than the HSE predictions. These results demonstrate the ability to detect adsorption processes as well as to control the electronic and magnetic properties of ZGNRs via chemical adsorption on surface anchoring sites.

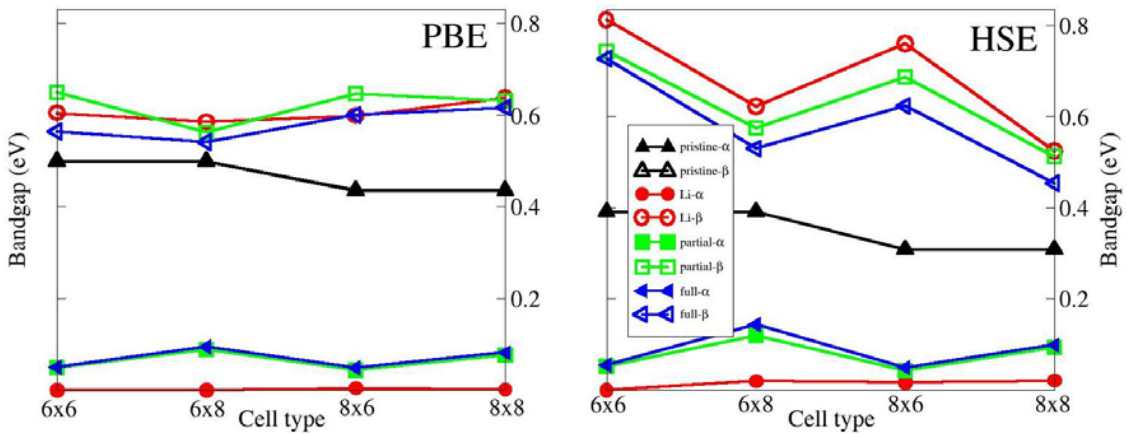

Figure 14: The influence of benzene adsorption on the α (full symbols) and β (open symbols) bandgaps of lithium doped ZGNRs. Presented is the dependence of the triplet spin-state bandgap on the unit-cell dimensions for hydrogen-terminated ZGNRs after lithium adsorption (red downward facing triangles), partial Li-benzene adsorption (green circles) and full Li-benzene adsorption (blue stars) as calculated using the PBE (left panel) and HSE (right panel) approximations. Open-shell singlet AFM results for the pristine nanoribbons appear as upward facing black triangles.

When examining the band-structure of the duplicated 8x8 ZGNR unit-cell (Fig. 15) one finds that the α spin bandgap appearing in the pristine ZGNRs (Fig. 15 (a)) considerably reduces upon lithium adsorption due to the appearance of flat mid-gap bands. Based on the partial DOS analysis these bands are attributed to lithium adatom states appearing in the vicinity of the Fermi energy of the system (Fig. 15 (b)). Consecutive adsorption of benzene molecules atop of lithium anchoring sites along either one or both faces of the ribbon results in bandgap increase accompanied by a character change from direct to indirect gap (Fig. 15 (c), (d), and (e)). The partial DOS diagrams indicate that, despite the fact that the benzene molecule states reside ~1.7 eV above the Fermi energy, the adsorption event alters the lithium states thus opening the bandgap of the system and changing its character. We note that apart from the mid-gap states, the general low-energy $\alpha$ electronic structure of the system is mostly preserved upon lithium and benzene adsorption. In Fig. 15 (f) a zoom in on the energy range in the vicinity of the $\beta$ spin Fermi energy is presented. Here, the direct bandgap of the pristine system turns indirect upon bare lithium atoms adsorption. Further adsorption of benzene molecules on the lithium linking sites somewhat reduces the bandgap via a relative rigid downshift of the conduction band minimum.

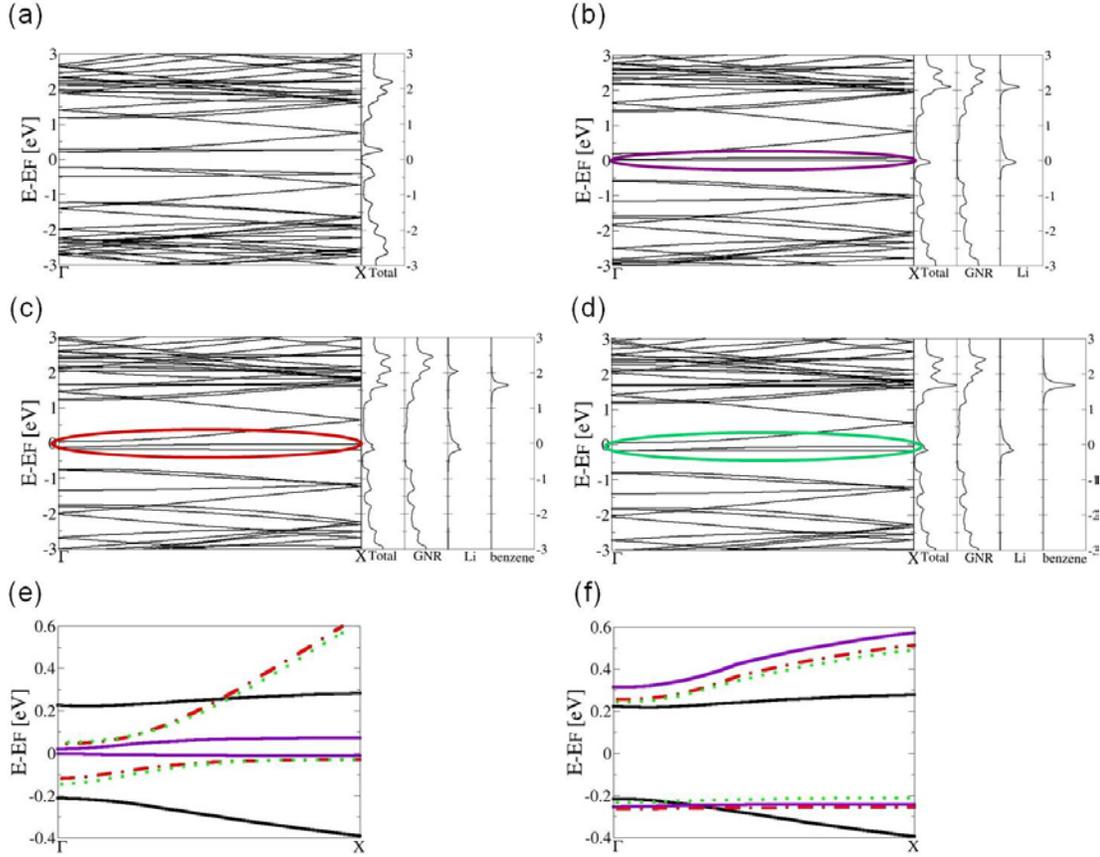

Figure 15: α spin band-structures, DOS, and PDOS of the 8x8 duplicated ZGNR unit-cell at the (a) open-shell singlet AFM pristine form; and the (b) triplet lithium adsorbed scheme; (c) Partial benzene adsorption scheme; and (d) Full benzene adsorption scheme. Panels (e) and (f) present a zoom in on the energy range of -0.4-0.6 eV around the Fermi energy of the α and β spin states, respectively, for the pristine (full black line), lithium adsorbed (full purple line), and partial (dashed red line) and full (green dotted line) Li-benzene adsorbed 8x8 ZGNR as calculated at the HSE/6-31G** level of theory. Fermi energies of all diagrams are set to zero.

To complete our understanding of the effects of lithium and benzene adsorption on the surface of GNRs we present in Fig. 16 a comparison of the total DOS of the pristine 6x7 AGNR (left panel) and α and β spin DOSs of the pristine 8x8 ZGNR (middle and right panels, respectively) with that of the chemically modified systems. The Fermi energies of the pristine GNRs are taken as the origin in all panels. When comparing the DOS of the pristine systems with their doped counterparts, it can be seen that the various adsorption schemes shift the Fermi energy of the AGNR and α spin states of the ZGNR towards higher energy values with increased DOS. We attribute this behavior to the charge transfer that occurs between the adsorbates and the underlying graphene surface. The β spin states exhibit an up-shift of both the valence band maximum and the conduction band minimum upon surface adsorption of lithium and benzene with an overall increased bandgap. These results further indicate that the electronic structure of GNRs is sensitive to surface molecular adsorption and may therefore be used to detect such events.

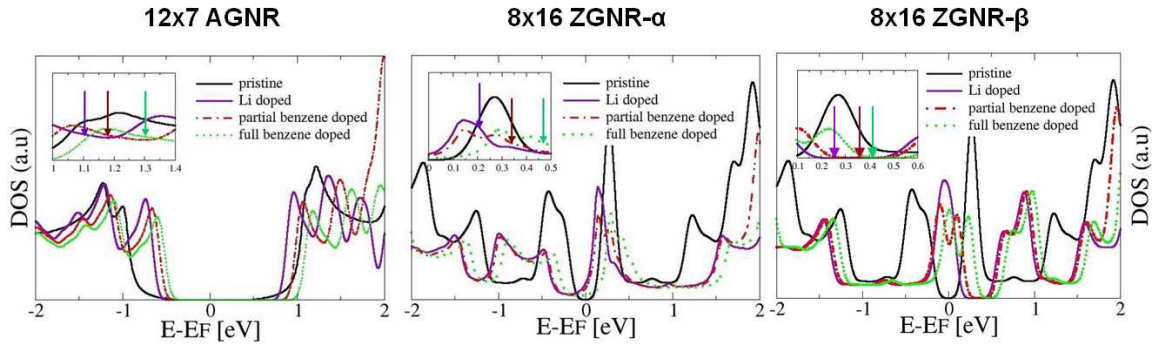

Figure 16: Comparison of the total DOS of pristine (full black line), lithium-doped (full purple line), benzene adsorption only on lithium sites above the surface (red dashed line) and benzene adsorption on lithium sites above and below the surface (green dotted line) of the 12x7 AGNR (left panel) as calculated at the HSE/6-31G** level of theory. Similar results for the α- and β- spin DOSs of the 8x16 ZGNR are presented in the middle and right panels, respectively. Fermi energies of the pristine GNRs are set as the origin of the energy axis. The corresponding Fermi energies of the respective chemically modified systems are indicated by the colored arrows in the insets.

Summary and conclusions

We have presented a comprehensive DFT analysis of lithium mediated benzene adsorption on two-dimensional graphene and both armchair and zigzag GNRs. When examining benzene adsorption on 2D lithium-graphene structures, the lithium atom encounters two π systems thus charge is drawn from the lithium atom to both the benzene molecule and the graphene surface while forming a bonded complex. Similar to bare lithium adsorption on 2D graphene, our results suggest that the ground state of the system is of closed-shell singlet nature with binding energies of the order of 1 eV/Li-benzene complex.

For the adsorption of benzene on lithium doped GNRs it was found that the favorable adsorption sites of lithium-benzene complexes are above hexagon centers near the edges of the ribbon. For all AGNRs studied a closed-shell ground state was obtained whereas for the zigzag systems considered the triplet spin state was found to be the most energetically stable. Bare lithium adsorption significantly reduced the bandgaps of GNRs, turning them either metallic in the case of AGNRs, or half-metallic in case of the ZGNRs, for sufficiently large adatom densities. The consecutive adsorption of benzene molecules on the lithium anchoring sites reopens a band-gap for AGNRs. This bandgap opening results mainly from up- and down-shifts of the low-lying conduction and valence bands, respectively, near the X-point while maintaining a direct character. In the case of ZGNRs, benzene adsorption alters the mid-gap $\alpha$ spin lithium bands changing the gap character from direct to indirect. For the $\beta$ spin states a rigid downshift of the conduction band minimum, upon benzene adsorption, results in reduction of the indirect gap characterizing the bare-lithium doped system. These results indicate that surface chemical adsorption may serve as a venue for controlling the electronic properties of GNRs and that the latter have potential to serve as sensing substrates in future miniaturized chemical detectors. Research aiming to estimate the sensitivity and susceptibility towards disturbing materials of such GNRs based chemical detectors is currently being pursued.

Acknowledgments: This work was supported by the Israel Science Foundation (ISF) under grant No. 1313/08, the European Community's Seventh Framework Programme FP7/2007–2013 under grant agreement No. 249225, the Center for Nanoscience and Nanotechnology at Tel-Aviv University, and the Lise Meitner-Minerva Center for Computational Quantum Chemistry.